\documentclass[epj,nopacs]{svjour}
\usepackage{graphicx,color}

\begin{document}
\title{The fine-tuning problem revisited \\ in the light of the Taylor-Lagrange renormalization scheme }
\subtitle{}
\author{P. Grang\'e \inst{1} \and J.-F.~Mathiot \inst{2} \and B. Mutet \inst{2} \and E. Werner \inst{3}}
\institute{Laboratoire de Physique Th\'eorique et Astroparticules, Universit\'e Montpellier II, CNRS/IN2P3, Place E. Bataillon, \\ F-34095 Montpellier Cedex 05, France
\and Clermont Universit\'e, Laboratoire de Physique
Corpusculaire,  BP10448, F-63000 Clermont-Ferrand, France
\and Institut f$\ddot u$r Theoretische Physik, Universit$\ddot a$t Regensburg, Universit$\ddot a$tstrasse 31, \ D-93053 Regensburg,  Germany}

\abstract{
We re-analyse the perturbative radiative corrections to the Higgs mass within the Standard Model in the light of the Taylor-Lagrange
renormalization scheme. This scheme naturally leads to completely finite corrections, depending on an arbitrary dimensionless scale. 
This formulation avoids very large individual corrections to the Higgs mass. In other words, it is a confirmation that the so-called fine-tuning problem in the Standard Model is just an artefact of the regularization scheme and should not lead to any physical interpretation in terms of the energy scale at which new physics should show up, nor to the appearance of a new symmetry. We analyse the characteristic physical  scales relevant for the description of these radiative corrections.
\PACS {
}}
\maketitle

\section{Introduction}
The experimental tests of the standard model of particle physics are entering a completely new era with the first $pp$ collisions at LHC (CERN) in the TeV energy range. On the theoretical side, any deviation from the theoretical predictions within the Standard Model will be a sign of new physics. From general arguments, new physics is expected to show up at some energy scale $\Lambda_{eff}$ above which the Standard Model should break down. We know that at most $\Lambda_{eff} < M_P$ where $M_P$ is the Planck mass ($\simeq 10^{19} GeV$) since gravitational effects become relevant, and cannot be neglected,  at that scale.

In any physical process, the theoretical consistency requirement of the Standard Model demands that the characteristic intrinsic momentum, denoted by  $\Lambda_k$, which is relevant for the description of this process  should be less or equal to $\Lambda_{eff}$.  Otherwise, new contributions of order $\left( \Lambda_k/\Lambda_{eff}\right)^n $ to the Lagrangian of the Standard Model should start to be sizeable. At tree level, this momentum is defined by the typical kinematical variables of the process, like $\sqrt{Q^2}$ in Deep Inelastic Scattering (DIS). It is thus under complete control. However, beyond tree level, one has to deal with internal momenta in loop contributions, that may be large. 

How large are they really? To answer this question, one has to enter into the renormalization procedure in order to define the physical amplitudes in terms of the bare ones calculated from the original Lagrangian of the Standard Model.

The renormalization procedures are of two kinds, depending on whether the renormalization of bare amplitudes is finite or infinite. The latter is the one which is widely used in standard perturbation theory {\it \`a la}  Feynman.
 In this scheme,  the choice of a regularization procedure is a necessary prerequisite to give a mathematical sense to a-priori divergent bare amplitudes. Two regularization methods are mainly used:
\begin{itemize}
\item The first one exhibits a very large mass scale, denoted by $\Lambda_C$. This mass scale is either a na\" ive cut-off in (Euclidean) four-momentum space, or the mass of Pauli-Villars (PV) particles in the PV regularization scheme. This explicit mass scale should be much larger then any characteristic energy, or momentum, scale relevant in the calculation of the theoretical physical amplitude.

\item The second one, the so-called dimensional regularization (DR) scheme, amounts to extending the space-time dimension $D$ away from $4$. The
divergences of the original amplitudes show up as singularities in $\varepsilon = 4-D$, with $\varepsilon > 0$. In this case, the bare amplitudes depend on a finite, and arbitrary, mass scale $\mu$.
\end{itemize}

The prototype of a finite renormalization scheme is the well-known BPHZ procedure. In this scheme, 
any bare amplitude is made finite by subtracting as many terms as necessary from the Taylor expansion at zero external momenta of the 
integrand. All Feynman integrals being convergent, no further regularization is required. 

In this study, we will focus on the Taylor-Lagrange renormalization scheme (TLRS) developped recently \cite{GW,grange}. This scheme  is based on the Epstein-Glaser \cite{EG} procedure to define physical amplitudes in terms of operator valued distributions (OPVD) acting on test functions. All amplitudes are then finite from the start, and depend on an arbitrary, finite, dimensionless scale.  The connection of the Epstein-Glaser formalism with the BPHZ 
renormalization scheme has been shown in Ref.~\cite{GB}. 

Beyond tree level, the first radiative corrections to the Higgs mass one has to consider in the Standard Model are radiative corrections  from a $t \bar t$ loop as well as Higgs and $W,Z$ bosons loops. Using a na\" ive cut-off  to regularize the bare amplitudes, these corrections immediately lead to a quadratic dependency of the regulated amplitudes on the cut-off scale $\Lambda_C$. The (square of the) physical mass, $M_H$, defined at the pole of the two-body Green's function, can be schematically written as
\begin{equation} \label{fine}
M_H^2 = M_0^2 + b \ \Lambda_C^2 + \ldots \ ,
\end{equation}
where $M_0$ is the bare mass of the Higgs particle, and $b$ is a combination of the top quark, $W,Z$ bosons and  Higgs masses. As it is, this equation has not much physical interest. It is just a definition of the bare mass as a function of the physical mass, in the spirit of the renormalization theory. 

The so-called fine-tuning problem arises if one wants to give some kind of physical interpretation to the bare mass $M_0$ \cite{kolda}. Since $\Lambda_C$ should be much larger than any characteristic energy scale relevant for the description of  the theoretical physical amplitude,  a large cancellation between $M_0^2$ and $b\,  \Lambda_C^2 $ should be enforced by hand --- hence the name fine-tuning --- unless $b$ is zero (the so-called Veltman condition \cite{velt}), or $M_H$ is very large.

Apart from the question of identifying the magnitude of $\Lambda_C$,  one may come back to the very origin of the fine-tuning problem, i.e. to the divergences of Feynman amplitudes in the standard approach. One way to tame these divergences is to rely on an extra symmetry, as for instance conformal symmetry \cite{conformal} or supersymmetry \cite{super}. However, to some extent, this is just a way to circumvent the problem but not to solve it! And the fine-tuning problem may in fact reappear in some way or another when these symmetries are explicitly broken \cite{barb}.

These divergences can be traced back to the violation of causality, originating from  ill defined products of distributions  at the same point \cite{Scharf,aste}. The correct mathematical treatment, known since a long time,  is to consider covariant fields
as OPVD, these distributions being applied on test functions with well-defined properties \cite{schwartz}. These considerations lead to the TLRS \cite{GW,grange}. 
 
 By construction, any bare amplitude calculated in TLRS is completely finite, and depends on an arbitrary, dimensionless, scale denoted by $\eta^2$. This scale is related to the shape of the test function near its boundaries in the ultraviolet (UV) or infrared (IR) regions. It can be traced back to the scale invariance of the UV and IR limits. Indeed, when any momentum $k$ tends to infinity or  $0$, $\eta^2 k$ tends also to infinity  or $0$. Since there are no extra large scales at all from TLRS, the fine-tuning problem in the standard model is absent, by construction, without the need to invoke any new symmetry. One would of course face the question of hierarchy if large intrinsic physical mass scales are present in the original Lagrangian ---  like in grand unified theories --- as in any other renormalization schemes. We shall not address this question in the present study.
  
 Our concern in the following is twofold. We shall first illustrate the above general remarks by the explicit calculation of the radiative corrections to the Higgs mass, in leading order of perturbation theory, and compare the results obtained in TLRS with standard procedures using a na\" ive cut-off or DR. Since the TLRS leads, as a by-product, to the BPHZ scheme, our conclusions do also apply to this scheme. The analysis of the fine-tuning problem within the BPHZ scheme has been mentioned already in Ref.~\cite{foda}, with similar conclusions. As compared to BPHZ, our scheme does not necessitate any ad-hoc subtraction order by order in perturbation theory. It also exhibits an explicit arbitrary scale which enables a renormalization group (RG) analysis of any physical amplitude.
 
Secondly, we shall analyse our results in terms of the characteristic scale $\Lambda_k$, and compare its value for both types, finite or infinite, of renormalization schemes. This may in turn have important consequences for the determination of the relevant momentum and/or energy scales at which new physics should show up.
 
 The plan of the article is the following. We recall in Sec.~\ref{general}  the general features of TLRS. We apply this scheme to radiative corrections to the Higgs mass in the Standard Model in Sec.~\ref{finetune}. We discuss our results in the light of the fine-tuning problem in Sec.~\ref{disc}, and draw our conclusions in Sec.~\ref{conc}.

\section{The Taylor-Lagrange renormalization scheme} \label{general}
Any  quantum field $\phi(x)$ - taken here as a scalar field for simplicity -
 should be considered as an OPVD \cite{schweber,collins,haag}. This has been known for a long time. However, its full significance for practical calculations was not fully recognized until recently \cite{GW,grange,EG,GB,Scharf,aste,Sto}. 
 
 As any distribution, quantum fields  should be defined by their application on test functions, 
 denoted by $\rho$, with well identified mathematical properties \cite{schwartz}. In flat space, the physical field ${\varphi}(x)$ is thus defined by \cite{GW}
\begin{equation} \label{conv}
\varphi(x) \equiv \int d^Dy \,\phi(y) \rho(x-y)\ ,
\end{equation}
in $D$ dimensions.
If we denote by $f$ the Fourier transform of the test function, we can further write ${\varphi}(x)$ in terms of creation and destruction operators, leading to
\begin{equation}
\!\varphi (x)\!=\!\!\int\!\frac{d^{D-1}{\bf p}}{(2\pi)^{D-1}}\frac{f(\varepsilon_p^2,{\bf p}^2)}{2\varepsilon_p}
\left[a^\dagger_{\bf p} e^{i{p.x}}+a_{\bf p}e^{-i{p.x}}\right],\ 
\end{equation}
with  $\varepsilon^2_p = {\bf p}^2+m^2$.  

From this decomposition, it is apparent that test functions should be attached to each fermion and boson fields. Each 
propagator being the contraction of two fields should be proportional to $f^2$. In order to have a dimensionless argument for  $f$, 
we shall introduce an arbitrary scale $\Lambda$ to "measure" all momenta. $\Lambda$ can be any of the masses of the constituents. 
To deal with massless theories, we shall consider  some a-priori arbitrary value. The final expression of any amplitude 
should be independent of $\Lambda$.

As recalled in Ref.~\cite{grange}, the test function $f$ should have three important properties:

{\it i)} the physical field $\varphi(x)$ in (\ref{conv}) should be independent of the choice of the test function $\rho$. This may be
achieved if $f$, the Fourier transform of $\rho$, is chosen 
among the partitions of unity (PU) (see \cite{grange} for more details on PU's). It is  a function of finite support which is $1$ everywhere except at the boundaries. This 
choice is also necessary in order to satisfy Poincar\'e invariance. This is due to the fact that if $f$ is a PU, any power of  $f$ is also a PU. In the limit where the test function goes to $1$ over the whole space, we then have $f^n \to f$ and Poincar\'e invariance is thus recovered.

{\it ii)} In order to be able to treat in a generic way singular distributions of any type, the test function is chosen as a super regular test function (SRTF). It is a function of finite
extension - or finite support -  vanishing, as well as all its derivatives, at its boundaries, in the UV and in the  IR domains. 

{\it iii)} The boundary conditions of the test function - which in this study is assumed  to depend on a  one dimensional variable $X$ - should embody a scale invariance inherent, in the UV domain for instance,  to the limit $X \to \infty$ since in this limit $\eta^2 X$ also goes
 to $\infty$, where $\eta^2$ is an arbitrary dimensionless scale. This can  be done by considering a
running boundary condition for the test function, i.e. a boundary condition which depends on the variable $X$ according to
\begin{equation} \label{running}
f(X \ge H(X)) = 0 \ \ \ \mbox{ for} \ \ \ \ H(X)\equiv \eta^2 X g(X)\ .
\end{equation}
This condition defines a maximal value, $X_{max}$, given by $f(X_{max})=0$.
In order to extend the test function to $1$ over the whole space, we shall consider a set of functions $g(X)$, denoted by $g_\alpha(X)$, where 
by construction $\alpha$ is a real positive number smaller than $1$. A typical
example of  $g_\alpha(X)$ is given in Ref.~\cite{grange}, where it is shown that in the limit $\alpha \to 1^-$, with $\eta^2>1$, the running support of the PU test function then
stretches over the whole integration domain, $X_{max} \to \infty$ and $f \to 1$. In this limit $g_\alpha(X) \to 1^-$.   This running condition is equivalent to having an ultra-soft cut-off \cite{grange}, i.e. an infinitesimal drop-off of the test function in the asymptotic limit, the rate of drop-off being governed by the arbitrary scale $\eta^2$. 
A similar scale invariance is also present in the IR domain, when $X \to 0$.

With these properties, the TLRS can be summarized as follows, first in the UV domain. Starting from a general amplitude $\cal A$ written for simplicity in a one dimensional space as
\begin{equation} \label{cala}
{\cal A} = \int_0^\infty dX \ T^>(X) \ f(X) \ ,
\end{equation}
where $T^>(X)$ is a singular distribution,
we apply the following general Lagrange formula to $f(X)$, after separating out an intrinsic scale $a$ from the (running) dynamical variable $X$
\begin{equation} \label{faX}
f^>(aX)=-\frac{X}{a^k k!}\int_a^\infty\frac{dt}{t} (a-t)^k \partial_X^{k+1}\!\left[ X^k f^>(Xt)\right].
\end{equation}
This Lagrange formula is valid for any order $k$, with $k>0$, since $f$ is chosen as a SRTF. It is therefore equal to its Taylor remainder for any $k$. After integration by part in (\ref{cala}), and using (\ref{faX}), we can thus express the amplitude $\cal A$ as
\begin{equation} \label{afin}
{\cal A} = \int_0^\infty dX \ \widetilde T^>(X) \ f(X) \ ,
\end{equation}
where $\widetilde T^>(X)$ is the so-called extension of the singular distribution $T^>(X)$. In the limit $f \to 1$,  it is given by \cite{GW}
\begin{equation} \label{Tex}
\widetilde T^>(X)\equiv\frac{(-X)^{k}}{a^k k!} \partial_X^{k+1} \left[ X T^>(X)\right] \int_a^{\eta^2} \frac{dt}{t} (a-t)^k \ .
\end{equation}
The value of $k$ in (\ref{Tex}) corresponds to the order of singularity of the original distribution $T^>(X)$ \cite{GW}. In practice, it can be chosen as the smallest integer, positive or null,  which leads to a non singular extension $\widetilde T^>(X)$.  If  in the absence of the test function $T^>(X)$ leads to a logarithmic divergence in (\ref{cala}), $k$ is $0$. It is $1$ if the divergence is quadratic.
With this choice for $k$, the extension of $T^>(X)$ is no longer singular due to the derivatives in (\ref{Tex}),  so that we can safely perform the limit $f \to 1$ in (\ref{afin}), and obtain
\begin{equation}
{\cal A} = \int_0^\infty dX \ \widetilde T^>(X) \ ,
\end{equation}
which is well defined but depends on the arbitrary dimensionless scale $\eta^2$. This scale is the only remnant of the presence of the test function.
Note that we do not need to know the explicit form of the test function in the derivation of the extended distribution $\widetilde T^>(X)$. 
We only rely on its mathematical properties  and on the running construction of the boundary conditions. 

The extension of singular distributions in the IR domain can be done similarly \cite{GW,grange}. For an homogeneous distribution in one
dimension, with $T^<[X/t]=t^{k+1} T^<(X)$, the extension of the distribution is given by
\begin{equation} \label{TIR2}
\widetilde T^<(X)=(-1)^{k}\partial_{X}^{k+1} \left[ \frac{X^{k+1}}{k!} T^<(X) \mbox{ln} \left(\tilde \eta X \right)\right] \ ,
\end{equation}
with $\tilde \eta = \eta^2-1$.
The usual singular distributions in the IR domain are of the form $T^<(X)=1/X^{k+1}$. In that case  $\widetilde T^<(X)$ reads
\begin{equation} \label{derlog}
\widetilde T^<(X)=\frac{(-1)^{k}}{k!} \partial_X^{k+1}\mbox{ln} \left(\tilde \eta X \right)\ ,
\end{equation}
where the derivative should be understood in the sense of distributions. Doing this, the extension $\widetilde T^<(X)$ is nothing else than the pseudo-function (Pf) of $1/X^{k+1}$ \cite{GW,grange,schwartz}
\begin{equation}
\widetilde T^<(X) = \mbox{Pf} \left( \frac{1}{X^{k+1}} \right) . 
\end{equation}
The extension $\widetilde T^<(X)$ differs from the original distribution 
$T^<(X)$ only at the $X=0$ singularity.

\section {Application to the fine-tuning problem} \label{finetune}
\begin{figure}[t]
\includegraphics[width=19pc]{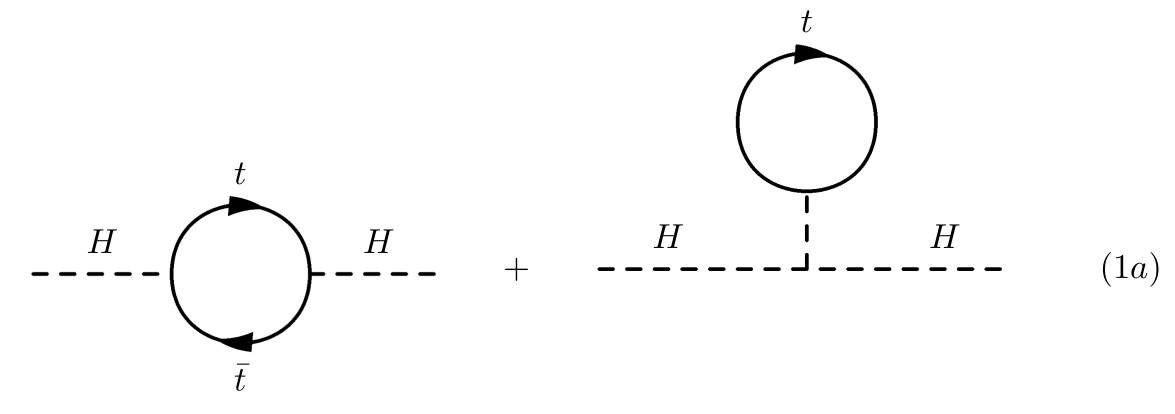}
\includegraphics[width=19pc]{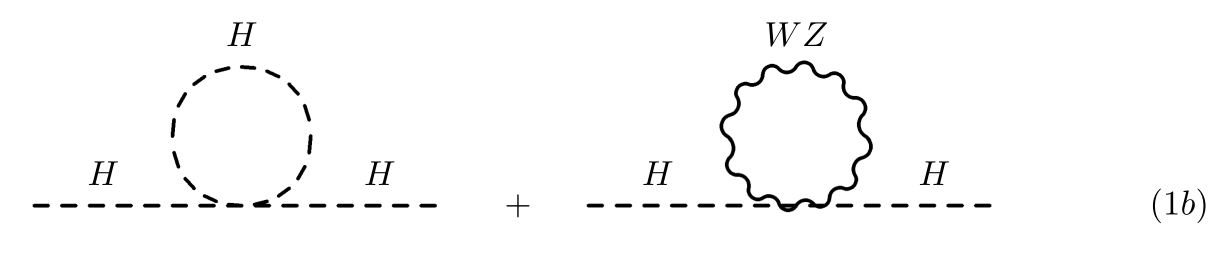}
\includegraphics[width=19.5pc]{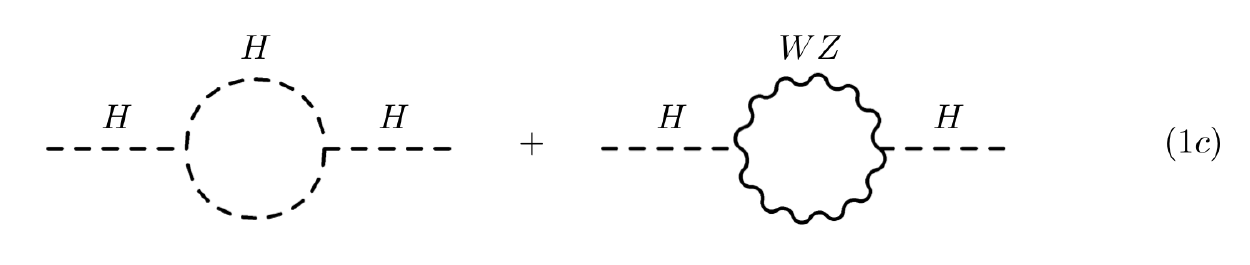}
\includegraphics[width=19.5pc]{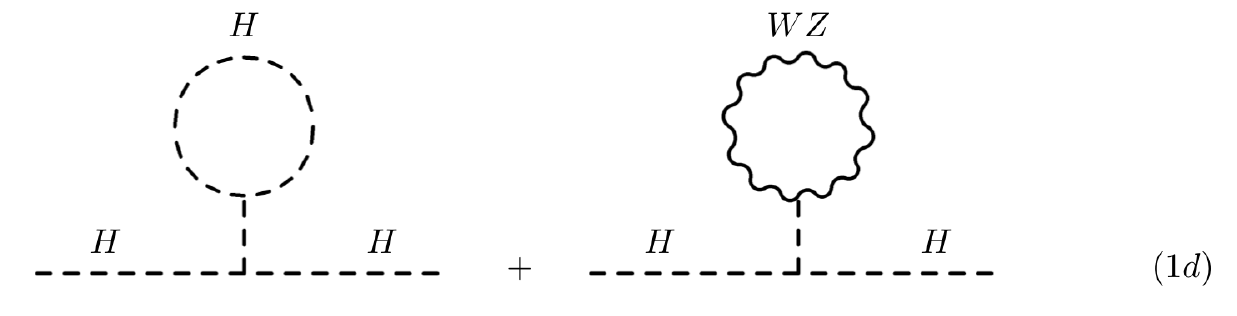}
\caption{Radiative corrections  to the Higgs mass in the Standard Model in second order of perturbation theory. For simplicity, we have not shown contributions from ghosts or Goldstone bosons.\label{standard}}
\end{figure}

In leading order of perturbation theory, the radiative corrections to the Higgs mass in the Standard Model are shown in Fig.~\ref{standard}. We have left out, for simplicity, all contributions coming from ghosts and Goldstone bosons. Each diagram in this figure gives a contribution to the self-energy $-i\Sigma(p^2)$, where $p$ is the four-momentum of the external particle, and we have
\begin{equation} \label{moh}
M_H^2=M_0^2+\Sigma(M_H^2)\ .
\end{equation}

Using a na\" ive cut-off to regularize the amplitudes, these radiative corrections lead to the well known mass correction
\begin{equation} \label{velt}
M_H^2 = M_0^2 + \frac{3 \Lambda_C^2}{8 \pi^2 v^2} \left[ M_H^2+2 M_W^2 + M_Z^2 -4 m_t^2\right] + \ldots \ ,
\end{equation}
where $m_t,M_{W,Z}$ and $M_H$ are the masses of the top quark, $W,Z$ and Higgs bosons respectively, and $v$ is the vacuum expectation value of the Higgs potential in the Standard Model. The dots include logarithmic corrections in $\Lambda_C$ as well as contributions independent of $\Lambda_C$ in the large $\Lambda_C$ limit. 

The calculation of the four different types of contributions shown in Fig.~\ref{standard} is very easy in TLRS. 
 Let us first illustrate the calculation of the simple Higgs loop contribution in Fig.~\ref{standard}.b.  In Euclidean space one has
\begin{equation} \label{f1b}
-i\Sigma_{1b,H}=-\frac{3iM_H^2}{2v^2}\int_0^\infty \frac{d^4k_E}{(2\pi)^4} \frac{1}{k^2_E+M_H^2} f\left( \frac{k^2_E}{\Lambda^2} \right) \ ,
\end{equation}
where $k^2_E$ is the square of the four-momentum $k$ in Euclidean space. As already mentioned in Sec.~\ref{general}, $\Lambda$ is an arbitrary momentum scale. The test function $f$ provides the necessary (ultra-soft) cut-off in the calculation of the integral.

After an evident change of variable, we get
\begin{eqnarray} \label{f1bred}
\Sigma_{1b,H}&=&\frac{3M_H^4}{32 \pi^2v^2} \int_0^\infty dX \frac{X}{X+1} f\left( \frac{M_H^2}{\Lambda^2}X \right)  \\ 
&=&\frac{3M_H^4}{32 \pi^2v^2} \int_0^\infty dX \left(1-\frac{1}{X+1}\right) f\left( \frac{M_H^2}{\Lambda^2}X \right) \nonumber\ .
\end{eqnarray}

The first term under the integral can be reduced to a pseudo-function, using (\ref{derlog}). Indeed, with $Z=1/X$, we have
\begin{eqnarray} \label{pf}
\int_0^\infty dX f(X) &=& \int_0^\infty \frac{dZ}{Z^2} f\left( \frac{M_H^2}{\Lambda^2} \frac{1}{Z} \right) \\
&=& \int_0^\infty dZ \ \mbox{Pf} \left(\frac{1}{Z^2} \right) \nonumber\\
&=& \left.-\frac{1}{Z}  \right|^\infty = 0  \nonumber \ .
\end{eqnarray}
The notation $f(u)\vert ^{a}$ simply indicates that $f(u)$ should be taken at the value $u=a$, the lower limit of integration being taken care of by the definition of the pseudo-function. This result is reminiscent of the property $\int d^D{\bf p} ({\bf p}^2)^n=0$, for any $n$, in DR \cite{collins}.

The self-energy thus writes
\begin{equation} \label{f2b}
\Sigma_{1b,H}=-\frac{3M_H^4}{32 \pi^2v^2} 
\int_0^\infty dX \frac{1}{X+1}f\left( \frac{M_H^2}{\Lambda^2}X \right) \ .
\end{equation}
The constant factor $M_H^2/\Lambda^2$ in the argument of the test function has no physical meaning since it can be absorbed by a rescaling of the arbitrary dimensionless scale $\eta^2$. This can be easily seen by applying the Lagrange formula (\ref{faX}) with the intrinsic scale $a=M_H^2/\Lambda^2$ and $k=0$. It can thus safely be removed \footnote{This could also be done more directly by choosing a particular value for $\Lambda$.}. 

We can now apply the Lagrange formula for $k=0$. Using the boundary condition on the support of the test function 
\begin{equation}
Xt \le H(X) = \eta^2 X g_\alpha(X) \ ,
\end{equation}
we finally get, in the limit $f \to 1$ 
\begin{eqnarray} \label{TLRS1}
\Sigma_{1b}&=&-\frac{3M_H^4}{32 \pi^2v^2} 
\int_0^\infty dX \partial_X \left( \frac{X}{X+1}\right) \int_1^{\eta^2} \frac{dt}{t} \nonumber \\
&=&- \frac{3M_H^4}{32 \pi^2 v^2} \mbox{ln}\left(\eta^2 \right)\ .
\end{eqnarray}
We shall come back in Sec.~\ref{disc} to the meaning of the limiting procedure $f \to 1$ in the presence of a physical cut-off $\Lambda_{eff}$ to define the domain of validity of the (effective) underlying theory.

It is easy to see that using a na\" ive cut-off on $k^2_E$ one would have obtained, in the large $\Lambda_C$ limit
\begin{equation}
\Sigma_{1b,H}^{C}=\frac{3M_H^2}{32 \pi^2v^2} 
\left[ \Lambda_C^2 - M_H^2\,\mbox{ln} \left(\frac{\Lambda_C^2}{M_H^2} \right)\right] \ .
\end{equation}
For completeness, we recall below the result of the direct calculation of (\ref{f1b}) in DR
\begin{equation} \label{DR1}
\Sigma_{1b,H}^{DR}=\frac{3M_H^4}{32 \pi^2v^2} 
\left[ -\frac{2}{\varepsilon} + c  - \mbox{ln} \left( \frac{\mu^2}{M_H^2} \right) \right] \ ,
\end{equation}
where $c=\gamma_E-1-\mbox{ln} 4\pi$ and $\gamma_E$ is the Euler constant.

We can proceed further to the calculation of the $\bar t t$ polarization correction indicated in Fig.~\ref{standard}.a (left diagram). This one depends explicitly on the (square of the) momentum $p$ of the external particle.
Similarly to (\ref{f1b}), in the Euclidean space, and using Feynman parametrization, one gets
\begin{eqnarray} \label{f1a}
-i\Sigma_{1a}(p^2)&=&\frac{12 i m_t^2}{v^2} \int_0^1 dx \int_0^\infty \frac{d^4k_E}{(2\pi)^4} \\
&&\times \frac{k_E^2-x(1-x)p_E^2-m_t^2}{\left[k^2_E+x(1-x)p_E^2+m_t^2\right]^2} f\left( \frac{k^2_E}{\Lambda^2} \right)\ , \nonumber
\end{eqnarray}
We thus have, 
\begin{eqnarray} \label{f1red}
\Sigma_{1a}(p^2)&=&-\frac{3  m_t^2}{4 \pi^2 v^2} \int_0^1 dx \ M^2(x,p^2)\int_0^\infty dX  \nonumber \\
&&\times \frac{X(X-1)}{(X+1)^2} f\left[ X \frac{M^2(x,p^2)}{\Lambda^2} \right] \ 
\end{eqnarray}
with $M^2(x,p^2)=x(1-x)p_E^2+m_t^2$ and  $X=k^2_E/M^2(x,p^2)$.
The integral over $X$, which we shall call $I$, can be decomposed into three parts, $I=I_1+I_2+I_3$, with
\begin{eqnarray}
I_1&=&\int_0^\infty  dX f\left[ X \frac{M^2(x,p^2)}{\Lambda^2} \right] \ ,\nonumber \\
\label {Ii} I_2&=&-3 \int_0^\infty  dX \frac{1}{X+1}f\left[ X \frac{M^2(x,p^2)}{\Lambda^2} \right] \ , \\
I_3&=&2 \int_0^\infty  dX \frac{1}{(X+1)^2}f\left[ X \frac{M^2(x,p^2)}{\Lambda^2} \right] \ .\nonumber \\
\end{eqnarray}
The first one  is zero according to (\ref{pf}). The second one can be calculated following the derivation of (\ref{f2b}), with an intrinsic scale $a=M^2(x,p^2)/\Lambda^2$ in the Lagrange formula (\ref{faX}) for $k=0$. This gives, using $\Lambda=M_H$,
\begin{equation}
I_2= \mbox{ln}\left[\eta^2\frac{M_H^2}{M^2(x,p^2)}\right]\ ,
\end{equation}
while the third one is trivial and gives $I_3=2$ in the limit $f\to 1$.

The self-energy correction from  the $\bar t t$ polarization diagram is thus
\begin{eqnarray} \label{TLRS2}
\Sigma_{1a}(p^2)&=&-\frac{3  m_t^2}{4 \pi^2 v^2} \int_0^1 dx \ M^2(x,p^2)\\
&&\times \left[-3\ \mbox{ln}\left[\eta^2\frac{M_H^2}{M^2(x,p^2)}\right]+2\right] \ . \nonumber
\end{eqnarray}

It is interesting to calculate directly (\ref{f1red}) without decomposing the integral $I$. To do that, we should apply the Lagrange formula for $k=1$. We thus get for this integral
\begin{equation} \label{direct}
I=- \int_0^\infty dX \ X \partial_X^2\left[
\frac{X^2(X-1)}{(X+1)^2}\right] \frac{1}{a} \int_a^{\eta^2} \frac{dt}{t}(a-t) \ .
\end{equation}
The self energy calculated in this way, and denoted by $\overline{ \Sigma}_{1a}(p^2)$, is given by
\begin{eqnarray} \label{TLRS3}
\overline{ \Sigma}_{1a}(p^2)&=&-\frac{3  m_t^2}{4 \pi^2 v^2} \int_0^1 dx\  M^2(x,p^2) \\
&&\times \left[\frac{3\eta^2 M_H^2}{\ M^2(x,p^2)}-3\ \mbox{ln}\left[\eta^2\frac{M_H^2}{M^2(x,p^2)}\right]-3\right]  \ . \nonumber
\end{eqnarray}
Comparing (\ref{TLRS2}) and (\ref{TLRS3}), we see that the calculation of the extension of a singular distribution is not unique. However, the self-energies differ either by a true constant (which depends on the arbitrary scale $\eta^2$, and is thus irrelevant in the calculation of the physical mass of the Higgs particle and more generally of any physical observable), or by a redefinition of the arbitrary scale $\eta^2$. They are thus said to be almost equivalent in the sense that they give identical physical, i.e. fully renormalized, amplitudes.

Using a na\" ive cut-off to calculate the self-energy (\ref{f1a}), one would have obtained 
\begin{eqnarray} \label{sigC}
\Sigma_{1a}^{C}(p^2)&=&-\frac{3 m_t^2}{4 \pi^2 v^2} \int_0^1 dx \ M^2(x,p^2)\\
&&\times \left[\frac{\Lambda_C^2}{\ M^2(x,p^2)}-3\ \mbox{ln}\left[\frac{\Lambda_C^2}{M^2(x,p^2)}\right]+2\ \right] \ . \nonumber
\end{eqnarray}
For completeness, we recall below the result in DR
\begin{eqnarray} \label{DR2}
\Sigma_{1a}^{DR}(p^2)&=&-\frac{3 m_t^2}{4 \pi^2 v^2} \int_0^1 dx \ M^2(x,p^2)\\
&&\times \left[-\frac{6}{\varepsilon} +3 c  -3\ \mbox{ln}\left[\frac{\mu^2}{M^2(x,p^2)}\right]+1\ \right] \ . \nonumber 
\end{eqnarray}

We can already see from these results that  TLRS and DR lead to a similar  $p^2$-dependent logarithmic term, with the identification $\eta^2=\mu^2/M_H^2$. They both depend on a completely arbitrary constant. The quadratic and logarithmic divergent terms using a cut-off procedure are transmuted in TLRS  into contributions depending on the arbitrary dimensionless scale $\eta^2$.

The other contributions to the radiative corrections to the Higgs mass indicated in Fig.~\ref{standard} can be calculated similarly. 
The final correction to the bare mass can thus be written schematically as
\begin{equation} \label{radfin}
M_H^2 = M_0^2 + \bar a\  \eta^2 + \bar b \  \mbox{ln}[\eta^2] + cte \ .
\end{equation}
This should be compared with the corrections indicated in Eqs.~(\ref{fine},\ref{velt}) when using a na\" ive cut-off. The exact expressions of $
\bar a$ and  $\bar b$ are of no physical interest since physical observables should be RG-independent of $\eta^2$. They are completely finite and  depend only on the masses of the top quark, $W,Z$ and Higgs bosons (and on the vacuum expectation value $v$ of the Higgs field). In fact, Eq.~(\ref{radfin}) legitimates  for the first time the use of perturbation theory at the level of the bare amplitude in contrast to Eqs.(\ref{sigC}) and (\ref{DR2}), which deal with perturbation theory on infinitely large corrections. 

Away from the on mass shell condition $p^2=M_H^2$, the constant term includes $p^2$-dependent logarithmic corrections which give rise to the well-known running of the mass. These corrections are identical in all three regularization schemes, as expected.

\section {Discussion} \label{disc}
We shall discuss our results in terms of the various scales appearing in the calculation of the radiative corrections to the Higgs mass, and more generally, to any physical observable. Three of them are of physical origin, and depend on the dynamical content of the underlying theory and on the kinematical conditions of the physical process under consideration, one is of mathematical origin, and two are completely arbitrary and are linked to the renormalization process.

The first physical scale is of course  $\Lambda_{eff}$ which defines the domain of validity of the underlying theory. It fixes, in the Standard Model, the energy scale above which new physics should show up. The second one corresponds to the kinematical scale defined by the physical process under consideration. It can be for instance $\sqrt{Q^2}$  in DIS. We shall call it $\Lambda_Q$ for simplicity. From a phenomenological point of view, in a bottom-up approach, $\Lambda_{eff}$ should correspond to the value of $\Lambda_Q$ for which theoretical predictions within the Standard Model are not corroborated by experimental results. 

The last physical scale we have already mentioned is the characteristic momentum relevant for the calculation of a given amplitude, called $\Lambda_k$. As we shall see below, it is intimately linked to the renormalization schemes which are used. Generally speaking, we should expect $\Lambda_Q < \Lambda_k$.

The mathematical scale is simply the cut-off $\Lambda_C$ used in the calculation of any integral over momenta in internal loops. It can not have any physical interpretation. It should be chosen sufficiently large, and one should check that any physical observable is independent --- within a given accuracy $\epsilon$ - of the exact value of $\Lambda_C$. In the literature, this mathematical scale is often taken equal to $\Lambda_{eff}$. We prefer here to separate clearly both scales since one has a physical interpretation while the other has not. This separation is indeed necessary from first principles since $\Lambda_{eff}$ could be infinite (for a renormalizable theory valid at all scales), while $\Lambda_C$ should be kept finite to regularize amplitudes in infinite renormalization schemes beyond tree level. As we shall explain below, this distinction is of particular interest when using finite renormalization schemes like TLRS.

The last two scales are related to the renormalization process. The first one is the arbitrary dimensionless scale $\eta^2$ introduced in (\ref{running})
in TLRS. It is the analog of the arbitrary mass scale $\mu$ of DR. The second one is the mass scale, called $R$, which is chosen to fix the bare parameters of the original Lagrangian in terms of physical measurable quantities\footnote{The scale $R$ may in fact correspond to a set of many different scales, if one chooses for instance to fix the masses and coupling constants at different momentum scales \cite{cheng}.}. It appears in both finite or infinite renormalization schemes. These two scales are closely related to the RG analysis, in the sense that all physical observables should be independent of both $\eta^2$ (or $\mu$) and $R$.

In order to determine $\Lambda_k$ from a quantitative point of view, we shall proceed in the following way. Contrarily to the common wisdom, the fact that the calculation of the self-energy necessitates a-priori a cut-off $\Lambda_C$ does not mean necessarily that all momenta up to $\Lambda_C$ are involved in the final value of the self-energy. Writing  the self-energy as
\begin{equation} \label{selfi}
\Sigma(p^2)=\int_0^{\Lambda_C^2}dk_E^2\  \sigma(k_E^2, p^2)\ ,
\end{equation}
we shall define the characteristic momentum $\Lambda_k$ by requiring that the reduced self-energy defined by
\begin{equation} \label{selfalpha}
\bar \Sigma(p^2)=\int_0^{\Lambda_k^2}dk_E^2 \ \sigma(k_E^2, p^2)
\end{equation}
differs from $\Sigma(p^2)$ by $\epsilon$ in relative value, i.e. with the constraint
\begin{equation} \label{eps}
\frac{\bar \Sigma(p^2)}{\Sigma(p^2)}=1-\epsilon\ ,
\end{equation}
provided we have $\vert \bar \Sigma(p^2) \vert < \vert \Sigma(p^2) \vert$.
In the Standard Model, $\epsilon$ can be taken of order $1 \%$. 

\begin{figure}[btph]
\includegraphics[width=20pc]{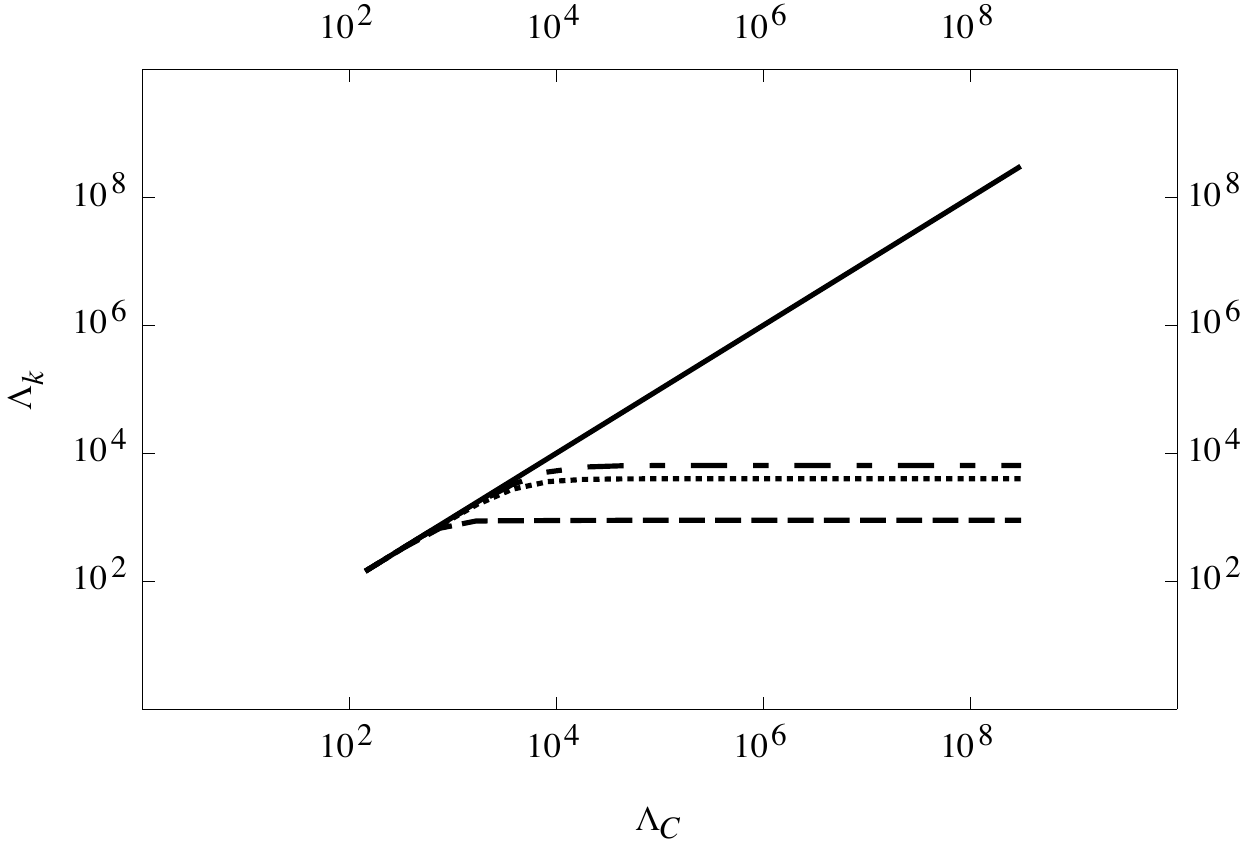}
\caption{Characteristic momentum scale $\Lambda_k$ calculated from the self-energy contribution $\bar \Sigma(M_H^2)$ defined  from
the condition (\ref{eps}), in two different regularization schemes:   with a na\" ive cut-off (solid line) and using TLRS (dashed line). The
calculation is done for $M_H=150$ GeV, with  $\eta^2=100$. We also show on this figure $\Lambda_k$ calculated
with the  fully renormalized self-energy (\ref{fullR}) for $p^2=-10 \ M_H^2$ (dotted line) and $p^2=-100 \ M_H^2$ (dash-dotted line).  \label{finescale}}
\end{figure}

We show in Fig.~\ref{finescale} the characteristic scale $\Lambda_k$ calculated for two typical  expressions of the self-energy of the Higgs particle, as a function of $\Lambda_C$. The first expression is the bare one given by $\Sigma(M_H^2)$ in (\ref{moh}), while the second one is the  fully (on-shell) renormalized amplitude, i.e. with both mass and wave function renormalization, defined by \cite{cheng}
\begin{equation} \label{fullR}
\Sigma_R(p^2)= \Sigma(p^2)- \Sigma(M_H^2)-(p^2-M_H^2) \left. \frac{d\Sigma(p^2)}{dp^2}\right |_{p^2=M_H^2}
\end{equation}
and calculated at two different values of $p^2$, $p^2=-10 \ M_H^2$ and $p^2=-100 \ M_H^2$. The calculation is done using (\ref{sigC}) for the cut-off regularization scheme, while (\ref{direct}) is taken for the calculation in TLRS, with an upper limit for the $X$ integral given by $\Lambda_C^2/M^2(x,p^2)$.

Note that the derivation summarized in Sec.~\ref{general} to calculate the extension of the distribution $T(X)$ in the UV domain is valid for any test function with finite support. Since $\widetilde T(X)$ in (\ref{afin}) is not singular anymore, all corrections from a finite support of $f$ in the upper limits of the integrals over $X$ or $t$ give a correction of order $1/X_{max}=M_H^2/\Lambda_C^2$.

The results indicated in Fig.~\ref{finescale} exhibit two very different behaviors. If one considers first the calculation of the bare amplitude, the use of a na\" ive cut-off regularization scheme does not allow to identify any characteristic momentum $\Lambda_k$. Since  $\Lambda_k$ is always very close to $\Lambda_C$, all momentum scales are  involved in the calculation of the bare self-energy. This is indeed a trivial consequence of the fact that the renormalization of the bare amplitude is infinite in that case. This would also be the case in any infinite renormalization scheme like DR. However, using TLRS, we can clearly identify a characteristic momentum $\Lambda_k$, since it reaches a constant value for  $\Lambda_C$ large enough. Note also that in this renormalization scheme, we can choose a value of $\Lambda_C$ which is arbitrary, as soon as it is much larger than any mass or external momentum of the constituents.  It can even be infinite, since it does not have any physical meaning, the only requirement being that physical amplitudes should be independent, within an accuracy $\epsilon$, of the precise value of $\Lambda_C$. This behavior is typical of finite renormalization schemes.

If we consider now the characteristic momentum scale relevant for the description of the fully renormalized amplitude $\Sigma_R$, we can also identify a finite value for $\Lambda_k$ since it saturates at sufficiently large values of $\Lambda_C$ compared to the typical masses and external momenta of the system. This behavior is extremely  similar to the result obtained in the above analysis of the bare amplitude $\Sigma$ using TLRS. This is again not surprising since the fully renormalized amplitude is also completely finite.  It depends only slightly on the external kinematical condition $\Lambda_Q$ (given here by $\sqrt{-p^2}$). In any case, the characteristic momentum scale  is of the order of $10$ times $\Lambda_Q$, and, what is more important, it is independent  of $\Lambda_C$. One can check  that $\Sigma_R$ is  of course identical in all renormalization schemes.

\section {Conclusions} \label{conc}
We have analysed in this article the fine-tuning problem in the Standard Model of particle physics in the light of the recently proposed Taylor-Lagrange regularization scheme. Since this scheme leads naturally to completely finite bare amplitudes - in contrast to a na\" ive cut-off regularization scheme which leads to quadratic divergences - we show explicitly that the fine-tuning problem is only an artefact of the regularization scheme, as one may already have suspected in the spirit of the renormalization theory.

In order to understand in more details the differences between the various regularization schemes, we have analysed the bare amplitudes, as well as the fully renormalized ones, in terms of the characteristic momentum relevant for the description of radiative corrections to the Higgs mass. In the case of the bare amplitudes, we find that this characteristic momentum is finite and independent of the cut-off $\Lambda_C$ when using TLRS, while it is as large as the cut-off scale for the cut-off regularization scheme. 
This forcludes any physical analysis in terms of a characteristic momentum when one uses a nat\" ive cut-off (or DR) on the bare amplitudes. The reason resides in an infinitely large bare amplitude for these latter regularization
schemes. 

On the contrary,  we can clearly identify the characteristic relevant momentum when using TLRS. This momentum is finite, while the value of 
$\Lambda_C$ can be very large, independently of the precise value of $\Lambda_{eff}$. This characteristic  momentum $\Lambda_k$ is in that
case completely determined by the dynamics of the underlying theory, as it should, and not by the mathematical properties of an
ill-defined integral. It should only satisfy the consistency condition $\Lambda_k < \Lambda_{eff}$. This condition can be interpreted in two different ways. If $\Lambda_{eff}$ is known (top-down approach), one should {\it verify} this condition in order to check the consistency of the effective theory. If $\Lambda_{eff}$ is not known (bottom-up approach), one should use this condition to {\it induce} a lower limit on $\Lambda_{eff}$, given  by $\Lambda_k$. This is the case of the Standard Model.

As expected, the equivalence between the use of different regularization schemes  is restored  if one analyses the fully renormalized amplitudes. In that case, one finds that the characteristic momentum is equivalent for all three regularization schemes, and it is of the order of the masses, or external momenta, of the constituents of the system. 
 
We can thus clearly identify two different scales: the first one, $\Lambda_C$, should be very large. It is just here to give a mathematical meaning, if necessary, to an integral, and should not have any physical interpretation. In TLRS, it can even be infinite, independently of the value of $\Lambda_{eff}$. The true physical scale which governs the dynamics of the system is the one given by the characteristic momentum $\Lambda_k$ relevant for the description of the fully renormalized amplitude. This scale is equivalent in all regularization schemes, as it should, and depends only on the dynamical content of the theory. It can be much less than $\Lambda_C$, and should be smaller than $\Lambda_{eff}$ for the theory to be valid.

A  remarkable feature of TLRS is that the identification of this characteristic momentum  scale in the calculation of any amplitude can be done already at the level of the bare amplitude, in four physical space-time dimensions. This is at variance with both the usual DR or cut-off regularization procedures.  Note that  since all calculations are done in four dimensions, the TLRS is particularly suited for theories involving $\gamma_5$ couplings, and can be extended naturally to supersymmetric theories, in contrast to DR \cite{jack}. 

Our analysis of radiative corrections to the Higgs mass in the Standard Model has shown that the characteristic momentum is of the
order of $10$ times the typical mass scale given, on the mass shell, by the physical Higgs mass. Moreover, once the various relevant scales have been clearly
identified, one can not give any information on the energy/momentum scale at which new physics should show up since this characteristic
momentum is independent of $\Lambda_{eff}$. It also does not rely on any new symmetry.

\begin{acknowledgement}
We would like to acknowledge enlightning discussions with G. Moultaka and J. Orloff, and financial support from CNRS/IN2P3. E. Werner is
grateful to A. Falvard for his kind hospitality at the LPTA.
\end{acknowledgement}



\begin{thebibliography}{10}
\bibitem{GW}
P. Grang\'e and E. Werner, {\it Quantum fields as Operator Valued Distributions and Causality}, arXiv: math-ph/0612011
and Nucl. Phys. B, Proc. Supp. {\bf 161} (2006) 75.
\bibitem{grange}
P. Grang\'e, J.-F. Mathiot, B. Mutet, E. Werner, Phys. Rev. {\bf D80} (2009)105012 ; Phys. Rev. {\bf D82} (2010) 025012.
\bibitem{EG}
H. Epstein and V. Glaser, Ann. Inst. Henri Poincar\'e, {\bf XIX A} (1973) 211.
\bibitem{GB}
J.M. Gracia-Bondia, Math. Phys. Anal. Geom. {\bf 6} (2003) 59;\\
J.M. Gracia-Bondia and S. Lazzarini, J. Math. Phys. {\bf 44} (2003) 3863.
\bibitem{kolda}
C.F. Kolda and H. Murayama, JHEP 0007 (2000) 035.
\bibitem{velt}
M. Veltman, Acta Phys. Polon. {\bf B12} (1981) 437.
\bibitem{conformal}
K.A. Meissner and H. Nicolai, Phys. Lett. {\bf B660} (2008) 260.
\bibitem{super}
E. Witten, Nucl. Phys. {\bf B188} (1981) 513;\\
S. Dimopoulos and H. Georgi, Nucl. Phys. {\bf B193} (1981) 150;\\
N. Sakai, Z. Phys. {\bf C11} (1981) 153;\\
L. Girardello and M. Grisaru, Nucl. Phys. {\bf B194} (1982) 65;\\
R.K. Kaul, Phys. Lett. {\bf B109} (1982) 19;\\
R.K. Paul and P. Majumdar, Nucl. Phys. {\bf B109} (1982) 36.
\bibitem{barb}
R. Barbieri, G.F. Giudice, Nucl. Phys. {\bf B306} (1988) 63.
\bibitem{Scharf} 
G. Scharf, {\it Finite QED: the causal approach}, Springer Verlag (1995).
\bibitem{aste}
A. Aste, {\it Finite field theories and causality}, Proceedings of  the International Workshop "LC2008 Relativistic nuclear and particle physics", 2008, Mulhouse, France, PoS(LC2008)001, and references therein.
\bibitem{schwartz}
L. Schwartz, {\it Th\'eorie des distributions}, Hermann, Paris 1966.
\bibitem{foda}
O.E. Foda, Phys. Lett. {\bf 124B} (1983) 192.
\bibitem{schweber}
S.S. Schweber, ``{\it An Introduction to Relativistic Quantum Field Theory}'', Ed. Harper and Row (1964), p.721.
\bibitem{collins}
J. Collins, `` {\it Renormalization}'',  Ed. Cambridge University Press, (1984), p.4.
\bibitem{haag}
R. Haag, ``{\it Local Quantum Physics: Fields, Particules, Algebras}'', Texts and Monographs in Physics, Springer-Verlag, Berlin, Heidelberg, 
New York,(2nd Edition,1996).
\bibitem{Sto} R. Stora, {\it Lagrangian  field theory}, Proceedings of Les Houches Summer School, Session 21, 
C. DeWitt-Morette and C. Itzykson eds., Gordon and Breach (1973).
\bibitem{cheng}
T.-P. Cheng and L.-F. Li, {\it Gauge theory of elementary particle physics}, Clarendon Press, Oxford (1984).
\bibitem{jack}
I. Jack and D.R.T. Jones, in {\it Perspectives on Supersymmetry}, World Scientific, Ed. G. Kane.
\end{thebibliography}
\end{document}